\catcode`\@=11					



\font\fiverm=cmr5				
\font\fivemi=cmmi5				
\font\fivesy=cmsy5				
\font\fivebf=cmbx5				

\skewchar\fivemi='177
\skewchar\fivesy='60


\font\sixrm=cmr6				
\font\sixi=cmmi6				
\font\sixsy=cmsy6				
\font\sixbf=cmbx6				

\skewchar\sixi='177
\skewchar\sixsy='60


\font\sevenrm=cmr7				
\font\seveni=cmmi7				
\font\sevensy=cmsy7				
\font\sevenit=cmti7				
\font\sevenbf=cmbx7				

\skewchar\seveni='177
\skewchar\sevensy='60


\font\eightrm=cmr8				
\font\eighti=cmmi8				
\font\eightsy=cmsy8				
\font\eightit=cmti8				
\font\eightbf=cmbx8				

\skewchar\eighti='177
\skewchar\eightsy='60


\font\ninei=cmmi9
\font\ninesy=cmsy9

\skewchar\ninei='177
\skewchar\ninesy='60


\font\tenrm=cmr10				
\font\teni=cmmi10				
\font\tensy=cmsy10				
\font\tenex=cmex10				
\font\tenit=cmti10				
\font\tensl=cmsl10				
\font\tenbf=cmbx10				
\font\tentt=cmtt10				
\font\tenss=cmss10				
\font\tensc=cmcsc10				
\font\tenbi=cmmib10				

\skewchar\teni='177
\skewchar\tenbi='177
\skewchar\tensy='60

\def\tenpoint{\ifmmode\err@badsizechange\else
	\textfont0=\tenrm \scriptfont0=\sevenrm \scriptscriptfont0=\fiverm
	\textfont1=\teni  \scriptfont1=\seveni  \scriptscriptfont1=\fivemi
	\textfont2=\tensy \scriptfont2=\sevensy \scriptscriptfont2=\fivesy
	\textfont3=\tenex \scriptfont3=\tenex   \scriptscriptfont3=\tenex
	\textfont4=\tenit \scriptfont4=\sevenit \scriptscriptfont4=\sevenit
	\textfont5=\tensl
	\textfont6=\tenbf \scriptfont6=\sevenbf \scriptscriptfont6=\fivebf
	\textfont7=\tentt
	\textfont8=\tenbi \scriptfont8=\seveni  \scriptscriptfont8=\fivemi
	\def\rm{\tenrm\fam=0 }%
	\def\it{\tenit\fam=4 }%
	\def\sl{\tensl\fam=5 }%
	\def\bf{\tenbf\fam=6 }%
	\def\tt{\tentt\fam=7 }%
	\def\ss{\tenss}%
	\def\sc{\tensc}%
	\def\bmit{\fam=8 }%
	\rm\setparameters\setbaselines\fi}


\font\twelverm=cmr12				
\font\twelvei=cmmi12				
\font\twelvesy=cmsy10	scaled\magstep1		
\font\twelveex=cmex10	scaled\magstep1		
\font\twelveit=cmti12				
\font\twelvesl=cmsl12				
\font\twelvebf=cmbx12				
\font\twelvett=cmtt12				
\font\twelvess=cmss12				
\font\twelvesc=cmcsc10	scaled\magstep1		
\font\twelvebi=cmmib10	scaled\magstep1		

\skewchar\twelvei='177
\skewchar\twelvebi='177
\skewchar\twelvesy='60

\def\twelvepoint{\ifmmode\err@badsizechange\else
	\textfont0=\twelverm \scriptfont0=\eightrm \scriptscriptfont0=\sixrm
	\textfont1=\twelvei  \scriptfont1=\eighti  \scriptscriptfont1=\sixi
	\textfont2=\twelvesy \scriptfont2=\eightsy \scriptscriptfont2=\sixsy
	\textfont3=\twelveex \scriptfont3=\tenex   \scriptscriptfont3=\tenex
	\textfont4=\twelveit \scriptfont4=\eightit \scriptscriptfont4=\sevenit
	\textfont5=\twelvesl
	\textfont6=\twelvebf \scriptfont6=\eightbf \scriptscriptfont6=\sixbf
	\textfont7=\twelvett
	\textfont8=\twelvebi \scriptfont8=\eighti  \scriptscriptfont8=\sixi
	\def\rm{\twelverm\fam=0 }%
	\def\it{\twelveit\fam=4 }%
	\def\sl{\twelvesl\fam=5 }%
	\def\bf{\twelvebf\fam=6 }%
	\def\tt{\twelvett\fam=7 }%
	\def\ss{\twelvess}%
	\def\sc{\twelvesc}%
	\def\bmit{\fam=8 }%
	\rm\setparameters\setbaselines\fi}


\font\fourteenrm=cmr12	scaled\magstep1		
\font\fourteeni=cmmi12	scaled\magstep1		
\font\fourteensy=cmsy10	scaled\magstep2		
\font\fourteenex=cmex10	scaled\magstep2		
\font\fourteenit=cmti12	scaled\magstep1		
\font\fourteensl=cmsl12	scaled\magstep1		
\font\fourteenbf=cmbx12	scaled\magstep1		
\font\fourteentt=cmtt12	scaled\magstep1		
\font\fourteenss=cmss12	scaled\magstep1		
\font\fourteensc=cmcsc10 scaled\magstep2	
\font\fourteenbi=cmmib10 scaled\magstep2	

\skewchar\fourteeni='177
\skewchar\fourteenbi='177
\skewchar\fourteensy='60

\def\fourteenpoint{\ifmmode\err@badsizechange\else
	\textfont0=\fourteenrm \scriptfont0=\tenrm \scriptscriptfont0=\sevenrm
	\textfont1=\fourteeni  \scriptfont1=\teni  \scriptscriptfont1=\seveni
	\textfont2=\fourteensy \scriptfont2=\tensy \scriptscriptfont2=\sevensy
	\textfont3=\fourteenex \scriptfont3=\tenex \scriptscriptfont3=\tenex
	\textfont4=\fourteenit \scriptfont4=\tenit \scriptscriptfont4=\sevenit
	\textfont5=\fourteensl
	\textfont6=\fourteenbf \scriptfont6=\tenbf \scriptscriptfont6=\sevenbf
	\textfont7=\fourteentt
	\textfont8=\fourteenbi \scriptfont8=\tenbi \scriptscriptfont8=\seveni
	\def\rm{\fourteenrm\fam=0 }%
	\def\it{\fourteenit\fam=4 }%
	\def\sl{\fourteensl\fam=5 }%
	\def\bf{\fourteenbf\fam=6 }%
	\def\tt{\fourteentt\fam=7}%
	\def\ss{\fourteenss}%
	\def\sc{\fourteensc}%
	\def\bmit{\fam=8 }%
	\rm\setparameters\setbaselines\fi}


\font\seventeenrm=cmr10 scaled\magstep3		


\newdimen\rp@
\newcount\@basestretchnum
\newskip\@baseskip
\newskip\headskip
\newskip\footskip


\def\setparameters{\rp@=.1em
	\headskip=24\rp@
	\footskip=\headskip
	\delimitershortfall=5\rp@
	\nulldelimiterspace=1.2\rp@
	\scriptspace=0.5\rp@
	\abovedisplayskip=10\rp@ plus3\rp@ minus5\rp@
	\belowdisplayskip=10\rp@ plus3\rp@ minus5\rp@
	\abovedisplayshortskip=5\rp@ plus2\rp@ minus4\rp@
	\belowdisplayshortskip=10\rp@ plus3\rp@ minus5\rp@
	\normallineskip=\rp@
	\lineskip=\normallineskip
	\normallineskiplimit=0pt
	\lineskiplimit=\normallineskiplimit
	\jot=3\rp@
	\setbox0=\hbox{\the\textfont3 B}\p@renwd=\wd0
	\skip\footins=12\rp@ plus3\rp@ minus3\rp@
	\skip\topins=0pt plus0pt minus0pt}


\def\setbaselines{\maxdepth=4\rp@\baselinestretch=\@basestretchnum}


\def\baselinestretch{\afterassignment\@basestretch\@basestretchnum}
\def\@basestretch{%
	\@baseskip=12\rp@ \divide\@baseskip by1000
	\normalbaselineskip=\@basestretchnum\@baseskip
	\baselineskip=\normalbaselineskip
	\bigskipamount=\the\baselineskip
		plus.25\baselineskip minus.25\baselineskip
	\medskipamount=.5\baselineskip
		plus.125\baselineskip minus.125\baselineskip
	\smallskipamount=.25\baselineskip
		plus.0625\baselineskip minus.0625\baselineskip
	\setbox\strutbox=\hbox{\vrule height.708\baselineskip
		depth.292\baselineskip width0pt }}



\def\makeheadline{\vbox to0pt{\baselinestretch=1000
	\vskip-\headskip \vskip1.5pt
	\line{\vbox to\ht\strutbox{}\the\headline}\vss}\nointerlineskip}

\def\makefootline{\baselineskip=\footskip\line{\the\footline}}

\def\big#1{{\hbox{$\left#1\vbox to8.5\rp@ {}\right.\n@space$}}}
\def\Big#1{{\hbox{$\left#1\vbox to11.5\rp@ {}\right.\n@space$}}}
\def\bigg#1{{\hbox{$\left#1\vbox to14.5\rp@ {}\right.\n@space$}}}
\def\Bigg#1{{\hbox{$\left#1\vbox to17.5\rp@ {}\right.\n@space$}}}


\mathchardef\alpha="710B
\mathchardef\beta="710C
\mathchardef\gamma="710D
\mathchardef\delta="710E
\mathchardef\epsilon="710F
\mathchardef\zeta="7110
\mathchardef\eta="7111
\mathchardef\theta="7112
\mathchardef\iota="7113
\mathchardef\kappa="7114
\mathchardef\lambda="7115
\mathchardef\mu="7116
\mathchardef\nu="7117
\mathchardef\xi="7118
\mathchardef\pi="7119
\mathchardef\rho="711A
\mathchardef\sigma="711B
\mathchardef\tau="711C
\mathchardef\upsilon="711D
\mathchardef\phi="711E
\mathchardef\chi="711F
\mathchardef\psi="7120
\mathchardef\omega="7121
\mathchardef\varepsilon="7122
\mathchardef\vartheta="7123
\mathchardef\varpi="7124
\mathchardef\varrho="7125
\mathchardef\varsigma="7126
\mathchardef\varphi="7127
\mathchardef\imath="717B
\mathchardef\jmath="717C
\mathchardef\ell="7160
\mathchardef\wp="717D
\mathchardef\partial="7140
\mathchardef\flat="715B
\mathchardef\natural="715C
\mathchardef\sharp="715D


\def\err@badsizechange{%
	\immediate\write16{--> Size change not allowed in math mode, ignored}}

\baselinestretch=1000
\tenpoint

\catcode`\@=12					
\catcode`\@=11
\expandafter\ifx\csname @iasmacros\endcsname\relax
	\global\let\@iasmacros=\par
\else	\immediate\write16{}
	\immediate\write16{Warning:}
	\immediate\write16{You have tried to input iasmacros more than once.}
	\immediate\write16{}
	\endinput
\fi
\catcode`\@=12


\def\rmb{\seventeenrm}

\def\singlespace{\baselineskip=\normalbaselineskip}
\def\halfspace{\baselineskip=1.5\normalbaselineskip}
\def\doublespace{\baselineskip=2\normalbaselineskip}


\def\AB{\bigskip\parindent=40pt
        \centerline{\bf ABSTRACT}\medskip\halfspace\narrower}
\def\AE{\bigskip\nonarrower\doublespace}
\def\nonarrower{\advance\leftskip by-\parindent
	\advance\rightskip by-\parindent}


\def\boxit#1{\vbox{\hrule\hbox{\vrule\kern3pt
	\vbox{\kern3pt#1\kern3pt}\kern3pt\vrule}\hrule}}

\def\hence{\leavevmode\hbox{\bf .\raise5.5pt\hbox{.}.} }

\def\dalemb#1#2{{\vbox{\hrule height.#2pt
	\hbox{\vrule width.#2pt height#1pt \kern#1pt \vrule width.#2pt}
	\hrule height.#2pt}}}
\def\gtorder{\mathrel{\raise.3ex\hbox{$>$}\mkern-14mu
             \lower0.6ex\hbox{$\sim$}}}
\def\ltorder{\mathrel{\raise.3ex\hbox{$<$}\mkern-14mu
             \lower0.6ex\hbox{$\sim$}}}

\newdimen\fullhsize
\newbox\leftcolumn
\def\twoup{\hoffset=-.5in \voffset=-.25in
  \hsize=4.75in \fullhsize=10in \vsize=6.9in
  \def\fullline{\hbox to\fullhsize}
  \let\lr=L
  \output={\if L\lr
        \global\setbox\leftcolumn=\columnbox\global\let\lr=R \advancepageno
      \else \doubleformat \global\let\lr=L\fi
    \ifnum\outputpenalty>-20000 \else\dosupereject\fi}
  \def\doubleformat{\shipout\vbox{
    \fullline{\box\leftcolumn\hfil\columnbox}\advancepageno}}
  \def\columnbox{\leftline{\vbox{\makeheadline\pagebody\makefootline}}}
  \tolerance=1000 }
\twelvepoint
\doublespace
{\nopagenumbers{
\rightline{IASSNS-HEP-00/43}
\rightline{~~~June, 2000}
\bigskip\bigskip
\centerline{\rmb Remarks on a Proposed Super-Kamiokande Test for}
\centerline{\rmb Quantum Gravity Induced Decoherence Effects }
\medskip
\centerline{\it Stephen L. Adler
}
\centerline{\bf Institute for Advanced Study}
\centerline{\bf Princeton, NJ 08540}
\medskip
\bigskip\bigskip
\leftline{\it Send correspondence to:}
\medskip
{\singlespace\leftline{Stephen L. Adler}
\leftline{Institute for Advanced Study}
\leftline{Einstein Drive, Princeton, NJ 08540}
\leftline{Phone 609-734-8051; FAX 609-924-8399; email adler@ias.
edu}}
\bigskip\bigskip
}}
\vfill\eject
\pageno=2
\AB
Lisi, Marrone, and Montanino have recently proposed a test for quantum 
gravity induced decoherence effects in neutrino oscillations observed at 
Super-Kamiokande.  We comment here that their equations have the same 
qualitative form as the energy conserving objective state 
vector reduction equations discussed by a number of authors.   However, 
using the Planckian parameter value proposed to explain state vector 
reduction leads to a neutrino oscillation effect many orders of magnitude 
smaller than would be detectable at Super-Kamiokande.  Similar estimates hold 
for the
Ghirardi, Rimini, and Weber spontaneous localization approach to state 
vector reduction, and our remarks are relevant as well to proposed $K$ meson 
and $B$ meson tests of gravity induced decoherence.  
\AE
\bigskip\bigskip
\vfill\eject
\pageno=3
There has recently been considerable interest in testing for possible 
modifications in conventional quantum mechanics induced by Planck mass 
scale quantum fluctuations in the structure of spacetime.  In an effective 
field theory approach, these are   
plausibly argued [1] to have the form of an extra ``decoherence term''  
${\cal D}[\rho]$ in the standard density matrix evolution equation, which 
becomes 
$${d \rho \over dt}=-i[H,\rho] - {\cal D}[\rho]  ~~~,\eqno(1)$$
where $\rho$ is the density matrix, $H$ is the Hamiltonian, and the 
decoherence term ${\cal D}$ has the dimensions of energy.  

In many 
phenomenological applications, Eq.~(1) is specialized by making several    
additional assumptions about the structure of ${\cal D}$.  First of all, 
the theory of open quantum systems suggests that Eq.~(1) should correspond 
to the infinitesimal generator form  of a {\it completely positive} map [2] 
on $\rho$, which requires that ${\cal D}$  should have 
the Lindblad form [3] 
$${\cal D}[\rho]=\sum_n[\{\rho,D_n^{\dagger}D_n\}-2D_n \rho D_n^{\dagger}]
~~~.\eqno(2)$$
If one further requires the monotone increase of the von Neumann entropy 
$S=-{\rm Tr} \rho \log \rho$, and the conservation of energy, one adds the 
respective conditions that the ``Lindblads'' $D_n$ should be 
self-adjoint, $D_n=D_n^{\dagger}$, and that they should commute with 
the Hamiltonian, $[D_n, H]=0$.  One then arrives at the form 
$$\eqalign{
{\cal D}[\rho]=&\sum_n[D_n,[D_n,\rho]] ~~~,\cr
[D_n,H]=&0~,~~{\rm all}~~ n~~~.\cr
}\eqno(3)$$ 
Equation (3) is the starting point of an analysis recently given by 
Lisi, Marrone, and Montanino [4] of decoherence effects 
in the super-Kamiokande experiment, interpreted in terms of 
$\nu_{\mu}-\nu_{\tau}$ oscillations.

When specialized to a two-level quantum system, the only choices of 
$D_n$ that commute with $H$ are either $D_n=\kappa_n 1$, with 1 the unit 
operator, or $D_n=\lambda_n H$.  The first choice is evidently trivial, 
since it makes no contribution to Eq.~(3), and so can be ignored.  
Hence all terms in the sum over $n$ in Eq.~(3) have the same structure, 
corresponding to the second choice;  defining $\lambda^2=  
\sum_n \lambda_n^2$, it is no restriction to replace the sum on $n$ 
in Eq.~(3) with a single Lindblad $D=\lambda H$. In their 
analysis of the Super-Kamiokande data, Lisi et. al. define a parameter 
$\gamma$ by 
$$\gamma=2 {\rm Tr} \sum_nD_n^2=2  \lambda^2 {\rm Tr} H^2~~~,\eqno(4)$$
and deduce the bound 
$$\gamma < 3.5 \times 10^{-23} {\rm GeV}~~~.\eqno(5)$$ 
Since in the two-level neutrino system we have 
${\rm Tr} H^2={1 \over 2} k^2$, with 
$k=\Delta m^2/(2E)$, where $\Delta m^2=m_2^2-m_1^2$ is the neutrino squared 
mass difference and $E$ is the neutrino energy, the parameters $\lambda$  
and $\gamma$ are related by 
$$\lambda = {\gamma^{1\over 2}\over k} = 
{2 E \gamma^{1\over 2} \over \Delta m^2}
~~~.\eqno(6)$$
Thus, using the Super-Kamiokande [5] value $\Delta m^2=3 \times 10^{-3} 
{\rm eV}^2$, 
and their maximum neutrino energy of $E \sim 10^3 {\rm GeV}$, 
the bound of Eq.~(5) on $\gamma$ corresponds to a bound on $\lambda$ of 
$$\lambda < 4 \times 10^{12} {\rm GeV}^{-{1\over 2}}~~~.\eqno(7)$$

The possibility that there may be decohering modifications to the 
Schr\"odinger equation, or to the corresponding density matrix evolution 
equation, has been extensively discussed over the past twenty years in 
the context of models for objective state vector reduction.  As surveyed 
by Adler and Horwitz [6], the form of the density matrix evolution assumed 
in these discussions is the It\^o stochastic differential equation
$$d\rho=-i[H,\rho]dt-{1\over 8}\sigma^2[D,[D,\rho]]dt
+{1\over 2}\sigma [\rho,[\rho,D]]dW_t~~~,\eqno(8)$$
with $D$ a Hermitian Lindblad operator driving the decoherence, and
with $dW_t$ an It\^o stochastic differential obeying 
$$dW_t^2=dt~,~~dtdW_t=0~~~.\eqno(9)$$
(One can readily generalize Eq.~(8) to contain a sum over multiple Lindblads 
$D_n$, but this will not be needed in our analysis.)  Two differing choices 
of the Lindblad $D$ have been widely discussed in the literature.  The 
first [7], due to Ghirardi, Rimini, and Weber, as extended by Di\'osi
and by Ghirardi, Pearle, and Rimini, 
takes $D$ to be a localizing operator in coordinate space; we will discuss 
this case later on.  The second [8], emphasized recently by  
Percival and Hughston, 
takes $D$ to be the Hamiltonian $H$, and this is the case on which we shall 
focus.  As shown by Adler and Horwitz, when $D$ is taken to be the 
Hamiltonian, Eq.~(8) can be proved, with no approximations, 
to lead to state vector reduction to energy eigenstates with the correct 
probabilities as given by the quantum mechanical Born rule.  To account 
for the observed absence of macroscopic spatial superpositions, one 
has to invoke energy shifts associated with environmental interactions which 
differ for macroscopic objects at different spatial locations; whether this 
leads to an empirically viable model for state vector reduction is presently 
an open question.  

Making  the choice $D=H$ in Eq.~(8), and taking the stochastic expectation, 
leads to an evolution equation for the stochastic expectation of the 
density matrix identical in form with Eqs.~(1-3) used by Lisi et. al. in 
their Super-Kamiokande analysis.  If one assumes a quantum gravitational 
origin for the stochastic terms in Eq.~(7), then the natural estimate [9] for 
the parameter $\sigma$ is $\sigma \sim M_{\rm Planck}^{-{1\over 2}}$, which 
since $\sigma^2/8=\lambda^2$ corresponds to 
$$\lambda \sim (8 M_{\rm Planck})^{-{1\over 2}} \sim 10^{-10} 
{\rm GeV}^{-{1\over 2}}~~~,\eqno(10)$$
more than twenty orders of magnitude smaller than the Super-Kamiokande  
bound on $\lambda$.  The difference in magnitudes is so great that there 
is clearly no prospect of confronting the prediction of Eq.~(10) in 
the Super-Kamiokande experiment.  The discrepancy between 
this conclusion, and the much more optimistic one reached by Lisi et. al.,  
arises as follows.  Lisi et. al. assume, on the basis of the general form 
for the decoherence term given in Eq.~(1), the estimate 
$${\cal D} \sim  H^2/M_{\rm Planck}~~~,\eqno(11)$$ 
with $H$ a characteristic energy (such as the neutrino 
energy) of the system.  However, once the decoherence term is restricted 
to have the self-adjoint Lindblad form of Eq.~(3), which is explicitly 
assumed in the analysis of Lisi et. al., the double commutator 
structure implies that the estimate is changed to 
$${\cal D} \sim (\Delta H)^2/M_{\rm Planck}~~~,\eqno(12)$$
with $\Delta H=|E_1-E_2|$ the energy {\it difference} between the levels for a 
two-level system.  [Note that the estimate of
Eq.~(12) is manifestly independent of the zero point with 
respect to which 
energies $H$ are measured, whereas the estimate of Eq.~(11) is not.]  
If the energy difference 
arises from the mass difference between the two beam components, we evidently 
have $\Delta H \simeq \Delta m^2/(2E)=k$.  This gives the estimate  
$${\cal D} \sim {(\Delta m^2)^2  \over 4E^2 M_{\rm Planck}}~~~,\eqno(13)$$
which because of the small neutrino mass difference is much more pessimistic 
than that of Eq.~(11).  Analogous remarks apply to tests for gravitation 
induced decoherence effects in the $K$ and $B$ meson systems, with 
$\Delta m^2$ replaced by the appropriate squared mass difference.  

Although we have focused our discussion on the case of energy driven 
dissipation, because this corresponds to the analysis of Lisi et. al., 
if we assume instead the spontaneous localization model the estimates  
for the Super-Kamiokande experiment are equally pessimistic.  In the  
spontaneous localization model, $D$ in Eq.~(8) is taken (in the small 
separation approximation for single particle dynamics) as 
the coordinate operator $q$, and the 
parameter $\sigma^2$ is given by $\sigma^2=2\lambda\alpha$, 
with $\lambda=10^{-16} {\rm sec}^{-1}$ the localization frequency,   
and with ${\alpha}^{-{1\over 2}}=10^{-5} {\rm cm}$ the localization 
distance.  Thus in a two-level system, an estimate of the dissipative 
term ${\cal D}$ is 
$${\cal D} \sim  10^{-6} {\rm sec}^{-1} {\rm cm}^{-2} (\Delta q)^2
~~~;\eqno(14)$$
estimating $\Delta q=|q_1-q_2|$ as the separation between centers of the 
$\nu_{\mu}$ and $\nu_{\tau}$ wave packets resulting from their mass 
difference, we get 
$$\Delta q \sim {1 \over 2}\Delta m^2 L E^{-2}~~~,\eqno(15)$$ 
with $L$ the neutrino flight path.  To get an upper bound, we use the 
smallest Super-Kamiokande neutrino energy $E \sim 10^{-1} {\rm GeV}$ and 
the largest flight path $L \sim 10^4 {\rm km}$, giving 
$$\Delta q <  10^{-10} {\rm cm}~~~.\eqno(16)$$
When substituted into Eq.~(14) this gives the estimate 
$${\cal D} <  10^{-26} {\rm sec}^{-1} \sim 10^{-50} {\rm GeV}~~~,\eqno(17)$$
again much smaller than the limit  $\sim 10^{-23} {\rm GeV}$ placed on 
the decoherence term by the Super-Kamiokande experiment.  

To conclude, we see that in the spontaneous localization model for  
the decoherence term, as well as in the energy conserving  model, the 
predicted
effect for the Super-Kamiokande experiment is proportional to 
the square of the neutrino mass squared difference $\Delta m^2$, and 
hence on the basis of these models is unobservably small.  One could get an  
observable Super-Kamiokande effect within the framework of 
these models only by positing  
a much larger coefficient for the decoherence term than has  
generally been assumed in the state vector reduction context; 
it would be interesting to see if this has   
testable consequences elsewhere.

\centerline{\bf Acknowledgments}
This work was supported in part by the Department of Energy under
Grant \#DE--FG02--90ER40542.
\vfill\eject
\centerline{\bf References}
\bigskip
\noindent
[1] J. Ellis, J. S. Hagelin, D. V. Nanopoulos, and M. Srednicki,  
Nucl. Phys. B241, 381 (1984); J. Ellis, N. E. Mavromatis, and D. V. 
Nanopoulos, Mod. Phys. Lett. A12, 1759 (1997).  \hfill\break
\bigskip 
\noindent
[2] K. Kraus, Ann. Phys. (NY) 64, 311 (1971).\hfill\break
\bigskip
\noindent
[3] G. Lindblad, Commun. Math. Phys. 48, 119 (1976); V. Gorini, A. 
Kossakowski, and E. C. G. Sudarshan, J. Math. Phys. 17, 821 (1976); for a 
recent derivation, see S. L. Adler, Phys. Lett. A265, 58 (2000).  
\hfill\break
\bigskip
\noindent
[4] E. Lisi, A. Marrone, and D. Montanino, ``Probing quantum gravity  
effects in atmospheric neutrino oscillations'',  hep-ph/0002053.\hfill\break
\bigskip
\noindent
[5]  Super-Kamiokande Collaboration, Y. Fukuda et. al., Phys. Rev. Lett. 
81, 1562 (1998); Y. Totsuka, 
to appear in the Proceedings of {\it PANIC~}'99 (Uppsala, 
Sweden, 1999), 15th Particles and Nuclei International Conference;   
available at www-sk.icrr.u-tokyo.ac.jp/doc/sk/pub.\hfill\break  
\bigskip
\noindent
[6] S. L. Adler and L. P. Horwitz, J. Math. Phys. 41, 2485 (2000),
 quant-ph/9909026, 
J. Math. Phys. (in press).  For early references to the norm-preserving 
stochastic density matrix evolution, see N. Gisin, Phys. Rev. Lett. 52, 
1657 (1984); and L. Di\'osi, Phys. Lett. 129A, 419 (1988) and 132A, 
233 (1988). 
\hfill\break
\bigskip
\noindent
[7]  G. C. Ghirardi, A. Rimini, and T. Weber, Phys. Rev. D34, 470 (1986);L. 
Di\'osi, Ref. [6];
G. C. Ghirardi, P. Pearle, and A. Rimini, Phys. Rev. A42, 78 (1990).
\hfill\break
\bigskip
\noindent
[8]  I. C. Percival, Proc. R. Soc. Lond. A447, 189 (1994); 
L. P. Hughston,  
Proc. Roy. Soc. Lond. A452, 953 (1996); I. Percival, {\it Quantum 
State Diffusion}, Cambridge University Press, Cambridge, 1998.  
References to earlier literature 
can be found in Hughston's paper and Percival's book.\hfill\break  
\bigskip
\noindent
[9]  See, e.g., D. I. Fivel, quant-ph/9710042; L. P. Hughston, Ref. [8],   
and further references cited therein.  

\bye